\documentclass[seceq]{ptptex}

\usepackage{graphicx}
\usepackage{wrapft}

\title{
Simulating pasta phases by molecular dynamics and cold atoms 
--- Formation in supernovae and superfluid neutrons in neutron stars
}

\author{
Gentaro \textsc{Watanabe}$^{1,2}$%
}

\inst{
$^1$Asia Pacific Center for Theoretical Physics (APCTP)\\
POSTECH,
San 31, Hyoja-dong, Nam-gu, Pohang, Gyeongbuk 790-784, Korea\\
$^2$RIKEN, 2-1 Hirosawa, Wako, Saitama 351-0198, Japan
}

\abst{
In dense stars such as collapsing cores of supernovae and neutron
stars, nuclear ``pasta'' such as rod-like and slab-like nuclei are
speculated to exist.  However, whether or not they are actually formed
in supernova cores is still unclear.  Here we solve this problem by
demonstrating that a lattice of rod-like nuclei is formed from a bcc
lattice by compression. We also find that the formation process is
triggered by an attractive force between nearest neighbor nuclei,
which starts to act when their density profile overlaps, rather than
the fission instability.  We also discuss the connection between pasta
phases in neutron star crusts and ultracold Fermi gases.
}

\begin{document}

\maketitle


\section{Introduction}

Collapse driven supernova explosion, an explosion in the death of a
massive star, is one of the most dramatic phenomena in the universe
and has been a long-standing mystery in astrophysics \cite{colgate}.
One of the key ingredients to understand the mechanism of the
supernova explosion is the study of matter in the core of supernovae
\cite{bethe}.

State of matter in supernova cores also undergoes dramatic changes in
the process of the collapse.  Matter experiences an adiabatic
compression, in which the density in the central region of the core
increases from $\sim 10^{9}$ g cm$^{-3}$ in the beginning of the
collapse and finally reaches around the normal nuclear density $\sim
3\times 10^{14}$ g cm$^{-3}$ (corresponding to the number density of
nucleons $\rho_{0}=0.165$ fm$^{-3}$) just before the bounce.  It is
predicted that, in the final stage of the collapse, nuclei are
rod-like or slab-like rather than (roughly) spherical in the central
region of the core \cite{rpw,hashimoto}.  These non-spherical nuclei
are collectively called nuclear ``pasta''.

Recently, nuclear pasta attracts much attention of astrophysicists and
nuclear physicists [see, e.g., Refs.\ \citen{oyamatsu,nakazato}]
since it has been pointed out that the nuclear pasta 
can occupy 10 -- 20 $\%$ of the mass of supernova cores \cite{opacity}
and thus might have influences on the dynamics of supernova explosions
\cite{gentaro2,horowitz,opacity}.  However, a fundamental problem of
whether or not and how pasta phases are formed during the collapse is
still unclear.
In the present work \cite{qmd_formation}, we succeed
in numerically simulating the formation of a trianglar lattice of
rod-like nuclei from a bcc lattice of spherical nuclei in collapsing
cores and solve the above long-standing problem. In addition, we
discover that the formation process is very different from a generally
accepted scenario based on an instability with respect
to nuclear fission. Our work provides a solid basis for understanding
materials in supernova cores, which is indispensable for solving a
long-standing problem of the mechanism of supernova explosions.

\section{Fission Instability?}

In a generally accepted conjecture, based on the Bohr-Wheeler
condition for vanishing the fission barrier derived for isolated
nuclei, it is predicted that the formation of the pasta phases are
triggered by the fission instability with respect to the quadrupolar
deformation of spherical nuclei \cite{review}.  The essential point of
this prediction is that, at higher densities, the effect of the
Coulomb repulsion between protons in nuclei, which tends to make a
nucleus deform, becomes dominant over the effect of the surface
tension of the nuclei, which favors a spherical nucleus.  However,
Refs.\ \citen{brandt} and \citen{burvenich} have pointed out that the
background electrons, which have been ignored in the above prediction,
suppress the effect of the Coulomb repulsion between protons in nuclei
and the fission barrier never vanishes in the relevant density region.
These findings cast a doubt on the above conjecture based on the
fission instability.

\section{Theoretical Framework}

Since formation of the pasta phases is accompanied by dynamical and
drastic changes of the nuclear structure, an {\it ab-initio} approach
is called for.  The quantum molecular dynamics (QMD) \cite{aichelin},
which can properly incorporate the thermal fluctuations \cite{qmd_hot}
and enables us to simulate large systems with $\sim 10^3$ nucleons
or more \cite{qmd}, is a suitable approach for the present purpose.

We use the QMD Hamiltonian of Ref.\ \citen{maruyama}.  In our
simulations, we consider a system with protons, neutrons, and
charge-neutralizing background electrons in a cubic box with periodic
boundary condition.  Here, we shall focus on the case of the proton
fraction $x\simeq 0.39$ (total number of nucleons $N=3328$ with 1312
protons and 2016 neutrons).

We simulate the compression of the bcc phase of spherical nuclei in
the collapse [see Ref.\ \citen{qmd_formation} for detailed
procedures].  Starting from $\rho=0.15\rho_0$, we increase the density
by changing the box size $L$ (the particle positions are rescaled at
the same time).  Here the average rate of the compression is
$\lesssim\mathcal{O}(10^{-6})\ \rho_0/($fm$/c)$ yielding the time
scale of $\gtrsim 10^{5}$ fm$/c$ to reach the typical density region
of the phase with rod-like nuclei.  This is much larger than the time
scale of the change of nuclear shape and thus the dynamics observed in
our simulation is determined by the intrinsic physical
properties of the system.  We perform adiabatic compression and
isothermal compression at various temperatures [see Ref.\
\citen{qmd_formation} for details].  In all the cases, we observe the
formation of rod-like nuclei; here we show a typical example in
which we obtain a clear lattice structure of the rod-like nuclei.

\section{Results}

\begin{figure}[t]
\begin{center}
\resizebox{11.4cm}{!}
{\includegraphics{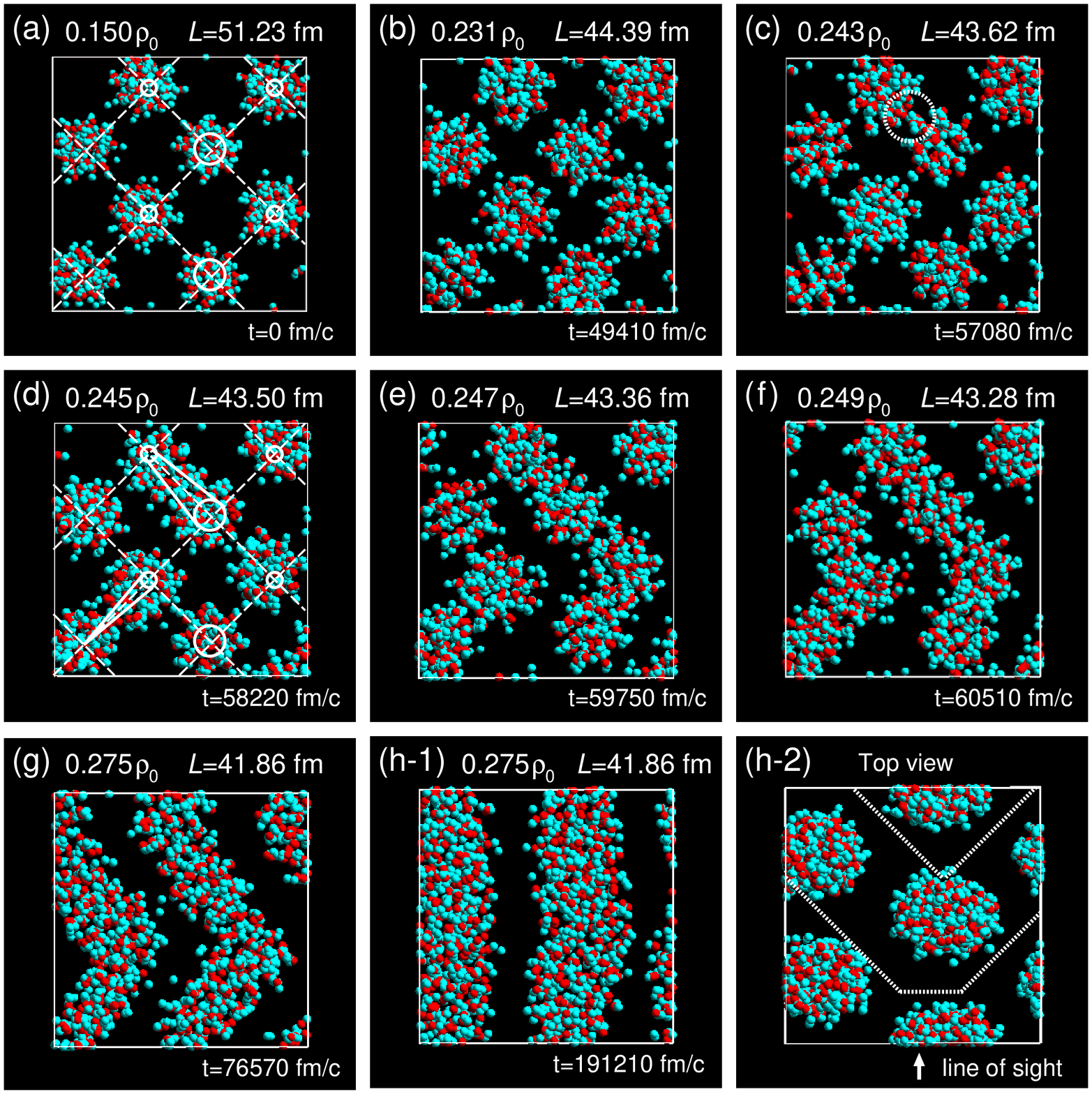}}
\caption{\label{snapshot}\quad Snapshots of the transition
process from the bcc lattice of spherical nuclei to the pasta phase
with rod-like nuclei.  The red particles show protons and the green
ones neutrons.  In panels (a)-(g) and (h-1), nucleons in a limited
region relevant for the two rod-like nuclei in the final state (h)
[surrounded by the dotted lines in panel (h-2)] are shown for
visibility.  The vertices of the dashed lines in panels (a) and (d)
show the equilibrium positions of nuclei in the bcc lattice and their
positions in the direction of the line of sight are indicated by the
size of the circles: vertices with a large circle, with a small
circle, and those without a circle are in the first, second, and third
lattice plane, respectively.  The solid lines in panel (d) represent
the direction of the two elongated nuclei: they take zigzag
configuration.  The box sizes are rescaled to be equal in the figures.
This figure is taken from Ref.\ \citen{qmd_formation}.
}
\end{center}
\end{figure}

Figure \ref{snapshot} shows the snapshots of the formation process of
the pasta phase in adiabatic compression.  Here, we start from initial
condition at $\rho=0.15 \rho_0$ and $T=0.25$ MeV ($t=0$ fm$/c$)
[Fig.\ \ref{snapshot}(a)].  At
$\rho\simeq 0.243 \rho_0$ [Fig.\ \ref{snapshot}(c)], the first pair of
two nearest-neighbor nuclei start to touch and fuse (dotted circle),
and then form an elongated nucleus [see, e.g., Fig.\
\ref{snapshot}(d)].  After multiple pairs of nuclei become such
elongated nuclei, we observe zigzag structure as shown in Fig.\
\ref{snapshot}(d).  Then these elongated nuclei stick together [see
Figs.\ \ref{snapshot}(e) and (f)], and all the nuclei fuse to form
rod-like nuclei as shown in Fig.\ \ref{snapshot}(g).  Finally, we
obtain a triangular lattice of rod-like nuclei after relaxation
[Figs.\ \ref{snapshot}(h-1) and (h-2)].

Note that before nuclei deform to be elongated due to the fission
instability, they stick together keeping their spherical shape [see
Fig.\ \ref{snapshot}(c)].  Besides, in the middle of the transition
process, pair of spherical nuclei get closer to fuse in a way such
that the resulting elongated nuclei take a zigzag configuration and
they further connect to form wavy rod-like nuclei.  This feature is
observed in all the other cases in which we obtain a clear lattice
structure of rod-like nuclei, and the above scenario of the transition
process is qualitatively the same also for those cases.  It is very
different from a generally accepted picture that all the nuclei
elongate in the same direction along the global axis of the resulting
rod-like nuclei and they join up to form straight rod-like nuclei
\cite{review}.

Instead of such a scenario, we have shown that the pasta phases are
formed in the following process [see Ref.\ \citen{qmd_formation} for
details].  When nearest neighbor nuclei are so close that the tails of
their density profile overlap with each other, net attractive
interaction between these nuclei starts to act.  This internuclear
attraction leads to the spontaneous breaking of the bcc lattice and
triggers the formation of the pasta phases.

\section{Simulating Pasta Phases by Ultracold Fermi Gases}

In the remaining part of this article, we shall discuss the connection
between the pasta phases and the ultracold atomic Fermi gases.  Pasta
phases can exist also in crusts of neutron stars.  There, pasta nuclei
are immersed in background electrons and a gas of dripped neutrons,
which is regarded to be in a superfluid state.

It has been pointed out that the low density neutron matter shows the
same universal property as unitary Fermi gases provided that the
interparticle separation $r_s$ between neutrons is much smaller than
the absolute value of the scattering length $a\simeq -18.5$ fm and is
much larger than the effective range $r_e\simeq 2.7$ fm
\cite{bertsch,baker} [see also Refs.\ \citen{gezerlis,abe,pethick}].
Here, ``universal'' means that properties do not depend on details of
the nuclear potential.

In dripped neutron gas in the pasta phases, the density is $\rho_n
=k_{\rm F}^3/(3\pi^2) \gtrsim 0.05$~fm$^{-3}$ and thus the
interparticle separation $r_s\gtrsim 1.7$ fm is comparable to $r_e$
(here, $k_{\rm F}$ is the Fermi wave vector of the ideal Fermi gas
with the same density).  However, Schwenk and Pethick
\cite{schwenk,pethick} have shown that in such a region of $k_{\rm
F}r_e\sim 1$, corrections from the universality in the equation of
state is accurately described within the effective range expansion and
no further details of nuclear forces are not needed.  The energy per
particle $E/N$ can be expressed in the same form as the universal case
but with a $r_e$-dependent factor $\xi(k_{\rm F}r_e)$: $E/N =
\xi(k_{\rm F}r_e)\ (3/5) E_{\rm F}$ with $E_{\rm F}\equiv
\hbar^2k_{\rm F}^2/2m$ and $m$ being the fermion (neutron, in the
present case) mass.  This means that the dripped neutron gas in the
pasta phases can be simulated by cold atomic Fermi gases with a narrow
Feshbach resonance.  At qualitative level, we can also expect that
unitary Fermi gases would provide a suggestive guidance for exploring
interesting physical effects related to the dripped superfluid
neutrons in the pasta phases.

In Refs.\ \citen{optlatunit} and \citen{vc}, we have studied
superfluid unitary Fermi gases in a one-dimensional (1D) periodic
potential $V_{\rm ext}(z)=s E_{\rm R} \sin^2{(q_{\rm B}z)}$, where $s$
is the laser intensity, $E_{\rm R}=\hbar^2q_{\rm B}^2/2m$ is the
recoil energy, $q_{\rm B}=\pi/d$ is the Bragg wave vector, and $d$ is
the lattice constant.  This setup resembles superfluid neutrons in the
pasta phase with slab-like nuclei, where $\rho_n\simeq 0.5\rho_0 \simeq
0.08$ fm$^{-3}$ and $d\simeq 15$ -- 20 fm and thus $E_{\rm F}/E_{\rm
R}=(k_{\rm F}/q_{\rm B})^2 \sim 40$ -- 70.  
The strength $V_0\equiv s E_{\rm R}$ of the optical lattice corresponds to the
depth of the bottom of the conduction band of neutrons 
measured from the average potential energy outside nuclei.
Since the neutron Fermi energy $E_{{\rm F}, n}$ for the density of
neutrons inside nuclei is $E_{\rm F,n}\sim 35$ -- 40 MeV and the
neutron chemical potential $\mu_n$ inside nuclei is $\mu_n \sim 10$ --
15 MeV, we estimate $V_0 \sim |\mu_n-E_{{\rm F}, n}| \sim 25$ MeV.  
Thus, we obtain $s\sim 25$ -- 50.

\begin{figure}[tbp]
\begin{center}\vspace{0.0cm}
\rotatebox{0}{
\resizebox{7.3cm}{!}{\includegraphics{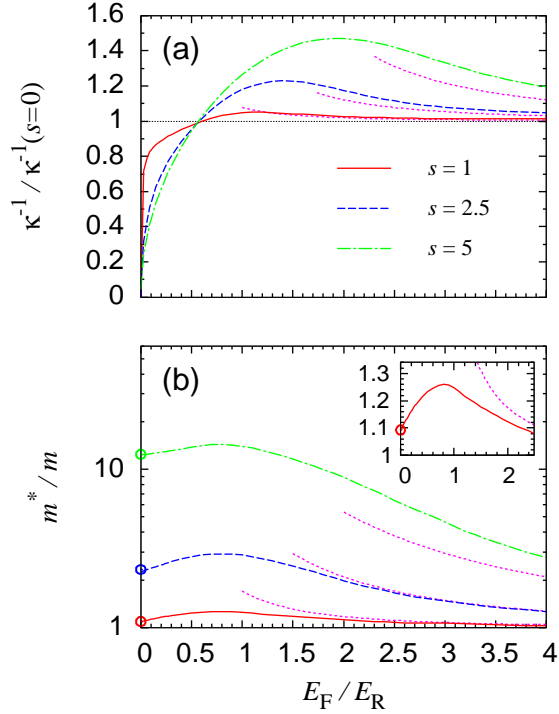}}}
\caption{\label{fig_kinv_meff_uni}(Color online)\quad 
Inverse compressibility $\kappa^{-1}$, and effective mass $m^*$ of the
unitary Fermi gas for $s=1$ (red), $2.5$ (blue), and $5$ (green).  The
solid, dashed, and dash-dotted lines show the results obtained by the
Bogoliubov-de Gennes (BdG) equations.  The dotted lines shows the
asymptotic expressions [Eqs.\ (\ref{expansionk-1}) and
(\ref{expansionm}) with $r_e=0$ and $\partial_n\xi=0$] obtained by the
hydrodynamic theory.  The $s=1$ results for $m^*$ are also shown in
the inset in the linear scale.  Adapted from Fig.\ (1) in Ref.\
\citen{optlatunit}.
}
\end{center}
\end{figure}

In Fig.\ \ref{fig_kinv_meff_uni}, we show the incompressibility 
$\kappa^{-1}=n\partial_n^2 e$ and the effective mass
$m^*=n(\partial_P^2 e)^{-1}$ of the unitary Fermi gas.
Here $e=e(n,P)$ is the energy density averaged over the unit cell 
as a function of the average (coarse-grained) density $n$ and
the quasimomentum $P$ of the superflow in the $z$ direction.

The rapid reduction of $\kappa^{-1}$ with decreasing $E_{\rm F}/E_{\rm
R}$ at small $E_{\rm F}/E_{\rm R}$ is due to the formation of bosonic
molecules induced by the periodic potential.  Maximum of $\kappa^{-1}$
and $m^*$ around $E_{\rm F}/E_{\rm R}\sim 1$ is the effect of the
energy band gap.  Since the tunneling rate through the barriers, which
relates to $m^*$, depends on the barrier height exponentially, $m^*$
increases drastically for larger $s$.  The enhancement of $m^*$
has been found also in dripped neutron gas in the crust of neutron
stars \cite{chamel}.

Taking account of the density dependence of $\xi$ in the case of the
large effective range, we derive $\kappa^{-1}$ and $m^*$ for $sE_{\rm
R}/E_{\rm F} \ll 1$ using the equation of state, $E/N = (3/5) \xi
E_{\rm F}$, within the hydrodynamic theory.  In the relevant region of
$k_{\rm F}r_e \gtrsim 3$, $\partial_n^2\xi$ is negligible, and keeping
up to the first order of $\partial_n\xi$ ($n\xi^{-1}\partial_n\xi
\lesssim 0.1$ in this region \cite{schwenk}), we obtain
\begin{equation}
\kappa^{-1} \simeq \frac{2}{3}\xi
E_{\rm F} \left\{ 1+\frac{1}{32} \xi^{-2} 
\left(\frac{sE_{\rm R}}{E_{\rm F}}\right)^2 
+ 3n\xi^{-1}\partial_n\xi 
\left[1+\frac{1}{32}\left(\frac{sE_{\rm R}}{E_{\rm F}}\right)^2\right]
\right\},
\label{expansionk-1}
\end{equation}
and
\begin{equation}
\frac{m^*}{m} \simeq 1+\frac{9}{32} \xi^{-2} 
\left(\frac{s E_{\rm R}}{E_{\rm F}}\right)^2 
\ .
\label{expansionm}
\end{equation}
Since $\xi$ is a monotonically increasing function of $k_{\rm F}r_e$
\cite{schwenk}, $m^*$ for non-zero $r_e$ is smaller than that of
unitary Fermi gases.

\section*{Acknowledgements}

Works reported in this paper have been done in collaborations with
H. Sonoda, K. Sato, K. Yasuoka, T. Ebisuzaki, F. Dalfovo, G. Orso,
F. Piazza, L. P. Pitaevskii, and S. Stringari.  We used MDGRAPE-2 and
-3 of the RIKEN Super Combined Cluster System, WIGLAF at the
University of Trento, and BEN at ECT$^*$.  We were supported by the
JSPS, by the MEXT through Grant No. 19104006, by EuroQUAM-FerMix, by
MiUR, by CNR, and by the EC Sixth Framework Programme.

%

\end{document}